\documentclass[useAMS,usenatbib]{mn2e}
\usepackage{epsfig,natbib}
\usepackage{amssymb}
\usepackage{graphicx}
\usepackage{placeins}
\usepackage{times}
\usepackage{epsfig}

\newcommand\ion[2]{#1{\sc #2}}

\title[On the sizes of $z\gtrsim2$ DLAs] {On the sizes of $z\gtrsim2$ Damped
Ly$\alpha$ Absorbing Galaxies\thanks{
Based on observations made with the Nordic Optical Telescope, operated
on the island of La Palma jointly by Denmark, Finland, Iceland,
Norway, and Sweden, in the Spanish Observatorio del Roque de los
Muchachos of the Instituto de Astrofisica de Canarias.
Based on observations collected at the European Organisation for Astronomical
Research in the Southern Hemisphere, Chile under programs
084.A-0303(A) and 086.A-0074(A).
}}
\author[J.-K. Krogager et al.]{J.-K. Krogager,$^{1}$\thanks{E-mail:
krogager@dark-cosmology.dk}
J. P. U. Fynbo$^{1}$,
P. M\o ller$^{2}$,
C. Ledoux$^{3}$,
P. Noterdaeme$^{4}$,
L. Christensen$^{1}$,\newauthor
B. Milvang-Jensen$^{1}$,
M. Sparre$^1$\\
$^{1}$Dark Cosmology Centre, Niels Bohr Institute, Copenhagen University, Juliane Maries Vej 30, 2100 Copenhagen O, Denmark\\
$^{2}$European Southern Observatory, Karl-Schwarzschild-Strasse 2, 85748 Garching bei M\"unchen, Germany\\
$^{3}$European Southern Observatory, Alonso de C\'ordova 3107, Vitacura, Casilla 19001, Santiago 19, Chile\\
$^{4}$UPMC-CNRS, UMR7095, Institut d'Astrophysique de Paris, F-75014 Paris, France\\
}

\begin{document}


\pagerange{1 -- 1} \pubyear{2011}

\maketitle


\begin{abstract}

Recently, the number of detected galaxy counterparts of $z \gtrsim 2$ Damped
Lyman-$\alpha$ Absorbers in QSO spectra has increased substantially so that we
today have a sample of 10 detections. M\o ller et al.\ in 2004 made the
prediction, based on a hint of a luminosity-metallicity relation for DLAs, that
\ion{H}{i} size should increase with increasing metallicity. In this paper we
investigate the distribution of impact parameter and metallicity that would result
from the correlation between galaxy size and metallicity. We compare our observations
with simulated data sets given the relation of size and metallicity.
The observed sample presented
here supports the metallicity-size prediction:
The present sample of DLA galaxies
is consistent with the model distribution. Our data also show a strong relation between impact parameter and
column density of \ion{H}{i}.
We furthermore compare the observations with several numerical simulations and demonstrate that the observations
support a scenario where the relation between size and metallicity is driven by feedback mechanisms
controlling the star-formation efficiency and outflow of enriched gas.

\end{abstract}


\section{Introduction}
The size of high redshift galaxies, its correlation with other properties like mass or luminosity,
and its evolution with redshift is a fundamental quantity to measure
\citep[e.g.,][]{ReesOstriker}. QSO absorption-line studies provide
an interesting way to address these relations as the method is sensitive to galaxies over
a large range of masses and sizes \citep[e.g.,][]{Nagamine07,Pontzen08,Cen10} 
and because it is complementary to the information
gathered from studying the stellar light of galaxies \citep[e.g.,][]{Ferguson}.

Damped Lyman-$\alpha$ Absorbers (DLAs) are believed to be the absorption
signature of neutral interstellar gas in galaxies (see \citet{Wolfe05} for a review).  When the first
sample of DLAs was collected it was found that the total cross-section for DLA
absorption was about five times higher than what would be expected from the
density and sizes of local spiral galaxies \citep[][see also M\o ller \& Warren
1998]{Wolfe86}. \citet{Wolfe86} and also \citet{Smith86} put forward the view that DLAs could be
forming disks with radii $\sim\sqrt{5}$ times larger than local spirals. In
this view galaxies were {\it larger} in the past, which would be at odds with
the prevailing hierarchical model of galaxy formation.  On the other hand,
\citet{Tyson1988} argued that DLAs could be caused by a population of many small
(much smaller than local spirals), gas-rich dwarf galaxies \citep[see also][]{Haehnelt98,  Ledoux98}.

Detecting the absorbing galaxies in emission is the most obvious way to determine the
sizes of DLAs from the distribution of impact parameters relative to the
background QSOs. Another method is to measure the extent of 21 cm or X-ray
absorption against extended background sources \citep{Briggs, Dijkstra}, but
such extended background sources are currently rare. Also, information on
sizes can be inferred from DLAs towards lensed QSOs \citep{Smette95,Ellison04,
Ellison07,Cooke10}, but such systems are also rare. Given the proximity of
the bright QSO to the line of sight to the absorbing galaxy the first approach
adopted was to search for Ly$\alpha$ emission from the absorbing galaxy as the
DLA itself here removes the light from the background QSO \citep{Foltz86}.
Two observing strategies
were adopted reflecting two different ideas about the nature of DLAs. Under
the assumption that the \ion{H}{i}-extent of DLA galaxies is small their impact parameters relative to the QSO
will also be small. Therefore a long-slit centered on the QSO should have a high
probability of also covering the DLA \citep[e.g.][and references
therein]{Hunstead90}. If on the other hand, DLAs arise from large \ion{H}{i} disks they may have
large impact parameters relative to the QSO and narrow-band imaging will be a
better approach as it eliminates the risk of missing the object on the slit
\citep{Smith86,Moller93}. Later, spectroscopy using integral field units
has also been attempted \citep{Christensen07,Peroux11, Bouche12}. 

In the period 1986--2010 only two galaxy counterparts of bona-fide intervening DLAs at
$z\gtrsim2$ ($\log{N_\mathrm{HI}/\mathrm{cm}^{-2}} > 20.3$, $z_{abs} \ll
z_{em}$ and $\Delta V > 6000\mathrm{km\,s}^{-1}$) were detected \citep[see][]{Moller04}
despite many more systems were observed.
This meager detection rate has gradually been
understood to reflect the fact that DLA galaxies, due to their cross-section selection,
trace the faint end of the galaxy luminosity function
\citep[][and references therein]{Fynbo08}.
\citet{Moller04} and \citet{Ledoux06}
found tentative evidence that DLA galaxies obey a metallicity-luminosity
relation similar to local galaxies and suggested that a
survey specifically targeting high metallicity DLAs should lead to a higher
success-rate in searches for galaxy counterparts. Indeed, this strategy has 
more than doubled the number of detections of $z\gtrsim2$ DLA galaxy 
counterparts in the literature in the last two years \citep[][]{Fynbo10,Fynbo11}.

The objective of this work is to test the correlation between metallicity and physical size
of DLA galaxies predicted by \citet{Moller04}. Thanks to the recent increase in number of detections
of emission from DLAs we can now analyze how the distribution of impact parameters and their
corresponding metallicities match up with the expected distribution from the model of \citet{Fynbo08},
based on the assumption of the previously mentioned correlation.

\section{Observations and Sample selection}
\label{obs}

Q\,0338$-$0005 was observed on November 9 2010 with the X-shooter
spectrograph mounted on UT2 of
ESOs Very Large Telescope following the same strategy as that described in
\citet{Fynbo10} and \citet{Fynbo11}, i.e.
securing 1 hr of integration at three position angles $0^\mathrm{o}$ and $\pm$60$^\mathrm{o}$.
For details on the reduction we refer to the description
in \citet{Fynbo10} and \citet{Fynbo11}.

\begin{figure}
	\includegraphics[width=0.48\textwidth]{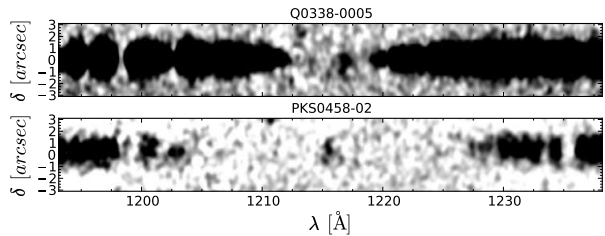}
	\caption{Two dimensional spectra from X-shooter of PKS0458$-$02 ({\it top})
	and Q0338$-$0005 ({\it bottom}) shifted to rest-frame wavelengths.
	In the centre of each spectrum the DLA troughs are seen with Ly$\alpha$ emission
	from the absorbing galaxies at spatial offsets of 0.31$\pm$0.04 arcsec
	and 0.49$\pm$0.12 arcsec, respectively. The spectra has been smoothed
	for illustrative purposes.
	\label{fig:newdla}}
\end{figure}

We detect the galaxy counterpart of the DLA in the trough of the damped
Ly$\alpha$ line in all three spectra (PAs of 0$^\mathrm{o}$, $\pm60^\mathrm{o}$).
The lower panel of fig.~\ref{fig:newdla} shows the stacked 
2D spectrum around Ly$\alpha$ where the Ly$\alpha$ emission line
can be seen in the bottom of the DLA trough.
Using the triangulation described in \citet{Moller04} and \citet{Fynbo10}
we infer an impact parameter of 0.49$\pm$0.12 arcsec. The metallicity is determined
to be [M/H]~$=-1.25\pm0.10$ based on low-ionization absorption lines from \ion{Si}{ii} 
in a high resolution UVES spectrum of the QSO (Ledoux, private communication)
and on the damped Ly$\alpha$ line from which we measure $\log{N_\mathrm{HI}/\mathrm{cm}^{-2}} 
= 21.05 \pm 0.05$.

The target PKS0458$-$02 was observed on November 24 2009 with X-shooter.
The slit was placed at a position angle of $-$60.4$^\mathrm{o}$
EofN with the purpose of measuring the precise impact parameter of the
$z=2.04$ DLA galaxy.
In the X-shooter spectrum (shown in the top panel of fig.~\ref{fig:newdla})
we measure an impact parameter of 0.31$\pm$0.04 arcsec.
We have combined the data from \citet{Moller04} with
our new measurement of the impact parameter.

The galaxy counterpart of the $z=2.58$ DLA towards Q\,0918$+$1636, 
reported in \citet{Fynbo11}, has here been observed in
a ground based imaging study using Point Spread Function (PSF) subtraction
to secure a more precise measure of the impact parameter.
The field was observed with the Nordic Optical Telescope (NOT) using the ALFOSC
instrument in service mode. 
The target was observed in the R-band over the nights March 27, 29 and April 9
2011, giving a total of 16\,380 sec exposure.
The images were reduced and combined using standard methods for CCD
imaging data.

\begin{figure}
	\includegraphics[width=0.48\textwidth]{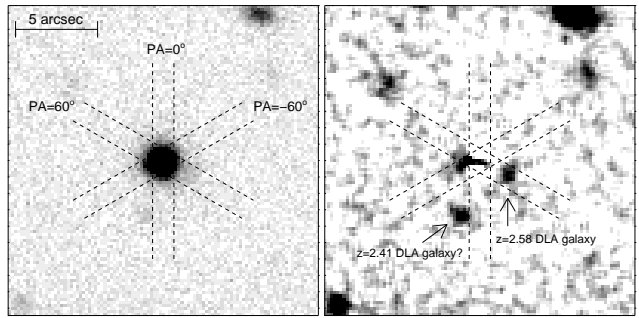}
	\caption{
	R-band image of the $20\times20$ arcsec$^2$ field of Q0918+1636.
	The left panel shows the image before
	PSF-subtraction and the right panel after PSF-subtraction and smoothing
	with a $3\times3$ pixel box-car filter. Also plotted are the three
	X-shooter slits used in the \citet{Fynbo11} study. The
	galaxy counterpart of the $z=2.58$ DLA detected in the X-shooter spectrum taken
	at $\mathrm{PA}=+60^\mathrm{o}$ is indicated with an arrow. Also seen is
	another galaxy at a PA of about 163$^\mathrm{o}$
	and an impact parameter of 3.5 arcsec.
	\label{fig:dlaim}}
\end{figure}

The field around Q0918$+$1636 is shown
in the left panel of Fig.~\ref{fig:dlaim}. The figure shows a 20$\times$20 arcsec$^2$
region from the combined NOT image. In the right panel we
show the field after PSF-subtraction and smoothing with a 3$\times$3 pixel
box-car filter. Two sources, both with $R$-band magnitudes of approximately 25,
are detected at impact parameters of 2.0 arcsec and 3.5 arcsec. Closer to the
position of the QSO there are residuals from the PSF subtraction. With dashed
lines we show the positions of the three X-shooter slit positions used by
\citet{Fynbo11}. As seen, the
continuum source at 2.0 arcsec seen in Fig.~\ref{fig:dlaim} coincides both in
position angle and impact parameter with the emission line source detected in
the X-shooter spectrum and must hence be the continuum counterpart of the
$z=2.58$ DLA galaxy. The source at 3.5 arcsec falls outside of the X-shooter
slits and it is on the basis of the data in hand not possible to establish if
this source may be related to the $z=2.41$ DLA also detected in the spectrum of
Q0918$+$1636 \citep{Fynbo11}.

\subsection{Sample selection}

In this paper we base our analysis on data combined from the literature along with our new data
for three systems of DLA galaxies; Q0338$-$0035, Q0918$+$1636 and PKS0458$-$02.
In addition to the systems in the literature we include a confirmed
candidate from an expanded sample based on the methods
in \citet{Christensen07} that has been confirmed by long-slit spectroscopy (Christensen et al., in preparation).
We note, however, that no other candidates from that work has been confirmed.
Bunker et al.\ in preparation also have a detection of Ly$\alpha$ emission from a DLA galaxy
using long-slit spectroscopy \citep[][their table 1]{Weatherley05}.
The redshift, metallicity and impact parameter of the systems used in our analysis are
shown in Table~\ref{DLAtable}.
The $z_{abs} \sim z_{em}$ DLA towards PKS0528$-$250 has been shown to be unrelated to the
nearby QSO and hence likely is similar to intervening DLAs
\citep{Moller93,Moller98}.

\section{The metallicity-size correlation}
\label{results}

\begin{table}
\caption{$z\gtrsim2$ DLAs with identified galaxy counterparts used in this 
study.\label{DLAtable}}
\begin{center}
\begin{tabular}{l c c c c}
\hline
\hline
QSO   & $z_\mathrm{abs}$ & log$(N_{\mathrm{H\,I}})$ & [M/H]  &  $b$\\
   	           &         &         &          & (arcsec) \\
\hline
Q2206$-$19$^{(1)}$		& 1.92 & 20.65$\pm$0.07 & $-$0.39$\pm$0.10\,$^{\mathrm{Zn}}$ & 0.99$\pm$0.05 \\
PKS0458$-$02$^{(1,5)}$ 	& 2.04 & 21.65$\pm$0.09 & $-$1.19$\pm$0.10\,$^{\mathrm{Zn}}$ & 0.31$\pm$0.04 \\
Q1135$-$0010$^{(7)}$    & 2.21 & 22.10$\pm$0.05 & $-$1.10$\pm$0.08\,$^{\mathrm{Zn}}$ & 0.10$\pm$0.01 \\
Q0338$-$0005$^{(5,8)}$  & 2.22 & 21.05$\pm$0.05 & $-$1.25$\pm$0.10\,$^{\mathrm{Si}}$ & 0.49$\pm$0.12 \\
Q2243$-$60$^{(4)}$		& 2.33 & 20.67$\pm$0.05 & $-$1.10$\pm$0.05\,$^{\mathrm{Zn}}$ & 2.80$\pm$0.20 \\
Q2222$-$0946$^{(2)}$ 	& 2.35 & 20.65$\pm$0.05 & $-$0.46$\pm$0.07\,$^{\mathrm{Zn}}$ & 0.8 $\pm$0.1  \\
Q0918$+$1636$^{(3,5)}$	& 2.58 & 20.96$\pm$0.05 & $-$0.12$\pm$0.05\,$^{\mathrm{Zn}}$ & 2.0 $\pm$0.1  \\
Q0139$-$0824$^{(6,9)}$  & 2.67 & 20.70$\pm$0.15 & $-$1.15$\pm$0.15\,$^{\mathrm{Si}}$ & 1.6 $\pm$0.05 \\
PKS0528$-$250$^{(1)}$	& 2.81 & 21.27$\pm$0.08 & $-$0.75$\pm$0.10\,$^{\mathrm{Si}}$ & 1.14$\pm$0.05 \\
Q0953$+$47$^{(10,11)}$	& 3.40 & 21.15$\pm$0.15 & $-$1.80$\pm$0.30\,$^{\mathrm{Si}}$ & 0.34$\pm$0.10 \\
\hline
\end{tabular}
\end{center}
References: (1) \citet{Moller04}; (2) \citet{Fynbo10}; (3) \citet{Fynbo11};
(4) \citet{Bouche12};
(5) this work;
(6) Christensen et al.\ in prep.; (7) \citet{Noterdaeme12};
(8) Ledoux, priv. comm.; (9) \citet{Wolfe08};
(10) Bunker, priv. comm.;
(11) \citet{Prochaska03}. \\
\end{table}

\begin{figure*}
	\includegraphics[width=0.98\textwidth]{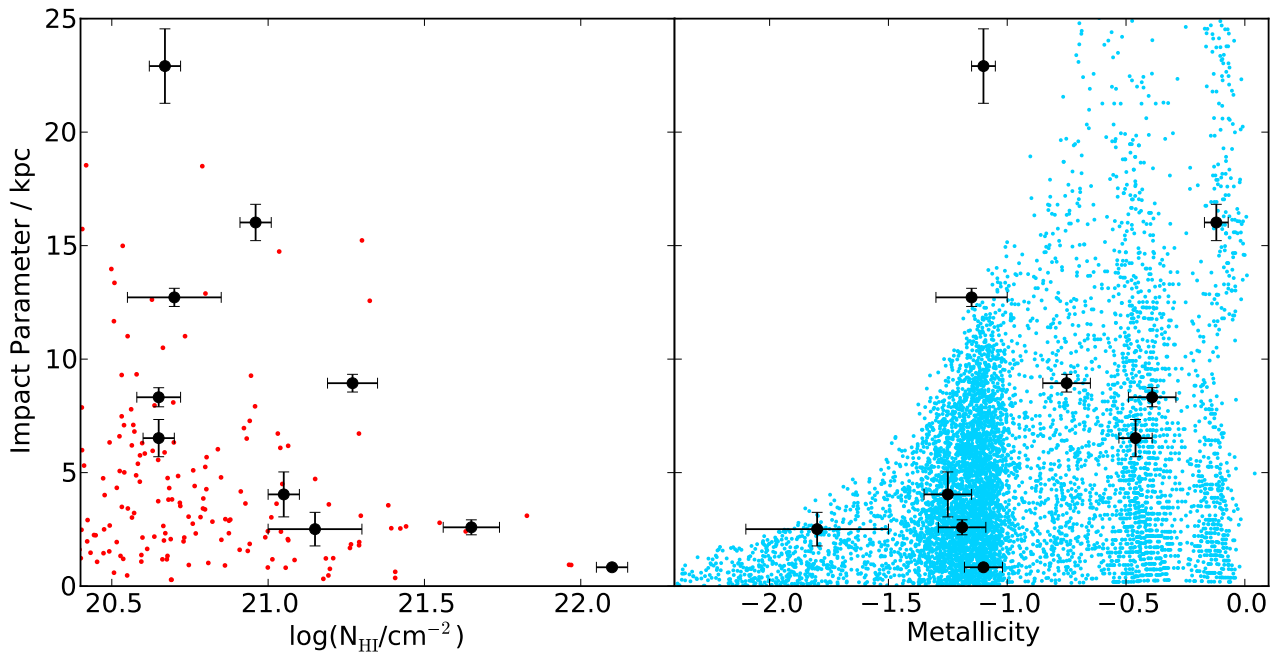}
	\caption{Impact parameters 
	plotted against \ion{H}{i} column density ({\it left}) and metallicity ({\it right}) for $z\gtrsim2$ DLAs 
	with securely identified galaxy counterparts, see table~\ref{DLAtable}. The observed impact parameters have been 
	converted to kpc at redshift $z=3$ instead of arcsec for easier comparison.
	The blue points in the right panel show the simulated distribution of impact parameters as a function of
	metallicity for DLA galaxies at $z=3$ from the model in \citet{Fynbo08}. The red points in the left panel
	show model points from \citet{Pontzen08}. 
	\label{fig:bZplot}}
\end{figure*}

\begin{figure}
	\includegraphics[width=0.48\textwidth]{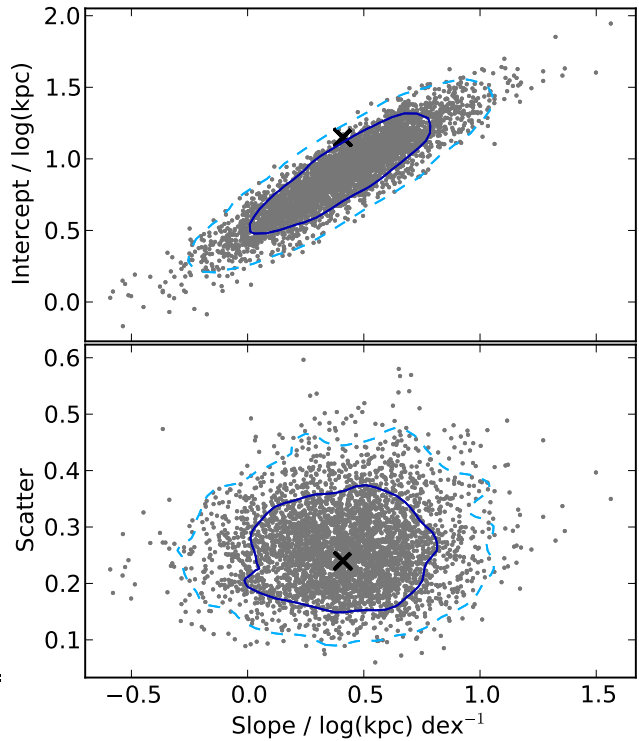}
	\caption{Distribution of ({\it top}) slope vs. intercept and ({\it bottom}) slope vs. scatter from the 4\,000 simulated
	data sets (grey points). The solid and dashed lines represent the 1$\sigma$
	and 2$\sigma$ confidence contours, respectively. The black cross shows
	the result from our fit to the data.
	\label{fig:hist}}
\end{figure}

The relation between metallicity and size
modeled by \citet{Fynbo08} is based on the assumption that DLAs seen in 
the spectra of background QSOs are caused by galaxies comparable to low 
luminosity Lyman Break Galaxies. In this model the galaxies are approximated by randomly 
inclined disks, each one assigned a size, metallicity and metallicity 
gradient based on its luminosity. The different relations between the 
quantities listed above are all based on local observationally 
motivated correlations for which indications of their validity at higher 
redshift have been seen. The more luminous galaxies have bigger disks, 
higher metallicities and shallower metallicity gradients. 
The luminosity $L$ is then drawn from the LBG luminosity function weighted by 
\ion{H}{i} cross section. An impact parameter is also assigned again with 
the luminosity dependent cross section of \ion{H}{i} as weighting factor. 
The random sight line is then given a metallicity from the before mentioned 
relations connecting [M/H] to $L$ and the radial dependence of 
metallicity. For further details see \citet{Fynbo08}.

\subsection{Testing the model}
Though the sample we present here is currently the largest at hand, we still only have ten points.
Therefore we need a simple statistical estimator to describe the data. We have chosen a two-parameter
fit of the form: $log(b) = \alpha + \beta \cdot $[M/H]. Since the internal scatter in the data is larger than the
individual measurement errors we fit a straight line to the measured data by least-squares minimization
without weights.

We then use the model described above to generate simulated data sets to which we can
compare the measured data. Because the measured data points are selected in different ways
we cannot simply draw ten random points from the model. In order to make a valid comparison,
we choose model points around the same metallicities as the measured points.

The simulated data points have metallicity taken randomly from a gaussian distribution around
each measured point using the measurement error as the width of the distribution.
We then take the distribution of impact parameters from the model at the given
metallicity and randomly pick an impact parameter.
We do this for all ten points to simulate a complete data set 4\,000 times. Every set of points is then fitted
by a straight line to give the intercept, slope and scatter.

The 4\,000 simulated data sets are shown in the right panel of fig.~\ref{fig:bZplot} as blue points along
with the measured data points in black. To make the plot less crowded only every fifth
simulated data set has been shown. The left panel of fig.~\ref{fig:bZplot} shows impact parameter
vs. neutral hydrogen column density for our data in big black points. The red small points in the panel
are results from \citet{Pontzen08}.

In fig.~\ref{fig:hist} we show the results of the fits to the data, both observed and simulated.
The figure shows the 1$\sigma$ ({\it solid}) and 2$\sigma$ ({\it dashed}) contours of the distribution
({\it grey points}) of fitted slope and intercept and slope and scatter.
The black cross in each panel point represents the best fitting values to the data:
$\alpha = 1.15 \pm 0.70 ~ \mathrm{log(kpc)} ,~~ \beta = 0.41 \pm 0.67 ~ \mathrm{log(kpc)\, dex}^{-1}$
with a scatter around the fit of $0.26$. The mean values of the intercept and slope
from the simulated data are
$0.90\pm0.28 ~ \mathrm{log(kpc)}$ and
$0.41\pm0.26 ~ \mathrm{log(kpc)\, dex}^{-1}$, respectively, with a mean scatter of 0.27.


\section{Discussion}
\label{discuss}

The sample of spectroscopically confirmed $z\gtrsim2$ DLA galaxies is now large 
enough to start determining statistical properties of DLA galaxies. Here
we have presented a new detection, namely of the DLA galaxy counterpart of the 
relatively metal poor DLA towards Q\,0338$-$0005 and have detected the $z=2.58$
DLA galaxy towards Q\,0918+1636 in the continuum.
Based on these and previous detections we have analyzed the distribution of impact parameters
vs. metallicity to study the underlying correlation between size and metallicity. The results presented
in fig.~\ref{fig:hist} show that our data are consistent with the model at the 1.2$\sigma$ level
for slope vs. intercept. Our data is well within the 1$\sigma$-contour in slope vs. scatter.
The agreement is also seen in the distribution shown in the right panel of fig.~\ref{fig:bZplot}.
There is, however, a single data point which is significantly outside
the model distribution (see fig.~\ref{fig:bZplot}: right panel, top part).
This can be explained as a result of the simplicity of the model. The correlations that go into
the model all have internal scatter. 
Including this scatter will result in a larger spread in physical size, hence making the upper edge of the
model distribution less tight, and thereby including the outlying point. 

An important point to note is that four of the data points have specifically 
been chosen to have high metallicity \citep{Fynbo10,Fynbo11}.
When comparing the observations with the model it is therefore important to remember
that the observed sample is biased towards high-metallicity DLA galaxies
that in our model are expected at the largest impact parameters.
At the same time we may be loosing points at the high-metallicity end because of 
dust bias \citep[e.g.,][]{Pei99,Khare2011} and due to the fact that our method using
three long-slits is less sensitive to high impact parameter DLAs, see \citet{Fynbo10}
for a detailed description of the method.
Nevertheless, the data on high-luminosity DLAs are consistent
with the picture in which gas at large impact parameters of tens of kpc cause DLA absorption. 
These galaxies seem to follow relations between size, luminosity
and metallicity comparable to the relations we see for disk galaxies in the local universe.

From hydrodynamic simulations it has also been investigated in which kind of
systems DLAs originate \citep[e.g.,][]{Cen10, Pontzen08, Razoumov06, Gardner01}.
In the simulation by \citet{Cen10}
DLAs at $z=3-4$ are dominated by low metallicity systems at large impact
parameters, typically $b=20-30$~kpc.  In this simulation the contribution from
gaseous disks to the DLA incidences is very small. Instead most DLAs arise in
filamentary gas.
The impact parameters at low metallicity from \citet{Cen10} are much larger than
what we observe. The only metal poor DLA is found at an impact parameter of
2.5 kpc, hence not supporting the model by \citet{Cen10}.
\citet{Razoumov06} also find a large number of DLAs at high impact parameter from
filamentary gas around large structures. But they find that the importance of these
filamentary DLAs decreases as they increase the resolution. 

\citet{Pontzen08} are able to reproduce the properties of DLAs well with their simulation,
which contains very detailed feedback mechanisms. From their simulation they
see positive correlations between the DLA cross-section and the mass of the
halo containing the DLA and between the metallicity and the halo mass (see
their fig.~4 and fig.~18). From the correlation between $\sigma_{\mathrm{DLA}}$ and
$M_{\mathrm{vir}}$ it is possible to infer a limit on the impact parameters of the
corresponding DLAs by assuming a spherical gas disk with a radius given by:
$R_{\mathrm{DLA}} \propto \sqrt{\sigma_{\mathrm{DLA}}}$. The largest possible cross-section found
in the most massive haloes is $\sigma_{\mathrm{max}} \approx 10^3\ \mathrm{kpc}^2$.
Hence, the impact parameters for DLAs around $z\approx3$ in their simulation
are expected to be $b \lesssim 30\mathrm{kpc}$.
The correlation of [M/H] and
halo mass suggests that the metal poor systems are to be found in smaller
haloes and therefore at smaller impact parameters. The opposite is true for
metal rich DLAs. This agrees well with the data presented here in
Fig.~\ref{fig:bZplot}.

We see a fairly strong anti-correlation between log($N_{\mathrm{HI}}$) and impact parameter
(Spearman rank = $-0.6$), see left panel in fig.~\ref{fig:bZplot}. A similar trend is seen
in studies by \citet{Moller98,Monier09} and by \citet{Peroux11}, however, we find a
tighter and steeper correlation in our data. \citet{Gardner01} also find a clear anti-correlation
between impact parameter and log(N$_{\mathrm{HI}}$) in their simulation of DLAs in different
cosmologies, including $\Lambda$CDM. They too find a shallower relation for DLAs at $z=3$.
This is most probably a result of the very simplified algorithms to handle cooling and feedback in their
simulation.
\citet{Pontzen08} find a clear relation between log(N$_{\mathrm{HI}}$) and $b$
({\it red points of} fig.~\ref{fig:bZplot}) that coincides very well with our observed distribution
as seen in the left panel of fig.~\ref{fig:bZplot}.
The good agreement between data and the simulation of \citet{Pontzen08}
indicates that the feedback mechanisms in their simulation that control
star formation and outflow of enriched gas are responsible for the correlation between
size and metallicity.

In conclusion, our analysis shows that our model well predicts
the statistical properties of the observed sample without
adjustment of any parameters, and as such provides strong
support for the underlying size-metallicity relation.
The nature of the metal poor population of DLA galaxies is, however,
still unknown, as these DLA counterparts are much more difficult to observe.

Furthermore, this sample of spectroscopically confirmed DLA galaxies serves as
constraints for future simulations that include modeling of DLAs.

\section*{Acknowledgments}

The Dark Cosmology Centre is funded by the DNRF. JPUF acknowledges support form
the ERC-StG grant EGGS-278202.
We also thank Andrew Pontzen for constructive discussion and Andrew
Bunker for sharing his data analysis with us.

\def\aj{AJ}
\def\araa{ARA\&A}
\def\apj{ApJ}
\def\apjl{ApJ}
\def\apjs{ApJS}
\def\apss{Ap\&SS}
\def\aap{A\&A}
\def\aapr{A\&A~Rev.}
\def\aaps{A\&AS}
\def\mnras{MNRAS}
\def\nat{Nature}
\def\pasp{PASP}
\def\aplett{Astrophys.~Lett.}

\bibliographystyle{mn}
\bibliography{thebib}

\end{document}